\documentclass{aastex631}

\newcommand{\nh}{$N_{\rm H}$}

\begin{document}

\title[An X-ray eclipse in V902~Mon]{An X-ray view of the Cataclysmic Variable V902~Mon: Discovery of an X-ray eclipse}
\author[0000-0002-2413-9301]{Nazma Islam}
\affil{Center for Space Science and Technology, University of Maryland, Baltimore County, 1000 Hilltop Circle, Baltimore, MD 21250, USA}
\affil{X-ray Astrophysics Laboratory, NASA Goddard Space Flight Center, Greenbelt, MD 20771, USA}

\author[0000-0002-8286-8094]{Koji Mukai}
\affil{Center for Space Science and Technology, University of Maryland, Baltimore County, 1000 Hilltop Circle, Baltimore, MD 21250, USA}
\affil{CRESST II and X-ray Astrophysics Laboratory, NASA Goddard Space Flight Center, Greenbelt, MD 20771, USA}

\correspondingauthor{Nazma Islam}
\email{nislam@umbc.edu}
\correspondingauthor{Koji Mukai}
\email{Koji.Mukai@umbc.edu}

\begin{abstract}

V902~Mon is one of a few eclipsing Intermediate Polars (IPs), and show deep eclipses in the optical lightcurves. The presence of a strong Fe K$\alpha$ fluorescence line in its X-ray spectrum and its low X-ray flux compared to other IPs suggests significant absorption, most likely from an accretion disk. In an observation carried out using the Nuclear Spectroscopic Telescope Array (NuSTAR), we confirm the presence of an X-ray eclipse in the energy resolved lightcurves, coincident with the optical AAVSO/CV-band lightcurves. Broadband X-ray spectral analysis using NuSTAR and XMM-Newton observations confirm a strong absorption \nh\ $\sim 10^{23}$ cm$^{-2}$ local to the source, along with a high equivalent width of about 0.7 keV for a Fe K$\alpha$ fluorescence line. We interpret this using a model similar to an Accretion Disk Corona source, which have a very high inclination and the compact object is heavily obscured by the body of the accretion disk. We propose that the primary X-rays from the accretion column in V902~Mon is hidden from our direct view at all times by the accretion disk. In this scenario, the observed scattered X-rays indicate substantial absorption of direct X-rays by the accretion disk. Additionally, a strong Fe fluorescence line suggests reprocessing of the radiation by a more extended region, such as the pre-shock region, which could be located a few white dwarf radii above the orbital plane.

\end{abstract}


\section{Introduction} \label{sec:intro}
Cataclysmic Variables (CVs) are semi-detached binaries consisting of a White Dwarf (WD) accreting matter from a low mass companion, which is mostly a late type main sequence star. A subclass of them are Intermediate Polars (IPs), where the accretion onto the WD is controlled by its magnetic fields ($B \sim 10^{6}-10^{7}$ G), forming truncated accretion disks and magnetically funnelling matter onto their poles from the inner edge of the truncated accretion disks. The accretion proceeds quasi-radially and form a strong shock with higher temperatures ($kT \sim$ 10--50 keV), and shock-heated plasma cools by emitting X-rays over a continuous temperature distribution, from the shock temperature to the WD photospheric temperature \citep{mukai2017}. The emergent X-rays from the shock heated multi-temperature plasma undergo absorption from a complex distribution of absorbers in the pre-shock region, where the  absorption \nh\ often exceeds 10$^{23}$ cm$^{-2}$ (\citealt{islam2021} and references therein). For typical IPs such as NY Lup and V1223 Sgr, the high specific mass accretion rates leads to the formation of shocks close to the WD surface, resulting in strong signatures of complex absorption and reflection signatures like a Compton hump and a strong Fe K$\alpha$ fluorescence line \citep{mukai2015}. Unlike Polars which are CVs with strong magnetic fields ($B \sim 10^{7}-10^{9}$ G) and the spin period of the WD is generally synchronized to the orbital period of the binary, the spin period of WDs in IPs are not synchronized to their orbital periods (for a review on IPs, see \citealt{patterson1994,hellier1996}). 
\par
V902~Mon (=IPHAS J062746.41+014811.3) is an eclipsing CV, which was discovered with the Isaac Newton Telescope (INT)/Wide Field Camera (WFC) Photometric H$\alpha$ Survey of the northern galactic plane \citep{witham2007}. It has an orbital period of 8.162 hours and a spin period of 2208 s was inferred from the optical photometry, and was therefore proposed to be an IP \citep{aungwerojwit2012}. The distance to the source from the Gaia EDR3 is $\sim$3.0 kpc, but with a considerable uncertainty (2.5--3.8 kpc) due to larger errors on the parallax measurements \citep{bailerjones2021}. In the optical lightcurves, it shows deep eclipses where the WD (whose photosphere is not directly observed) is inferred to be eclipsed by the secondary. This suggests that the system has a high inclination of about 82$^{\circ}$ \citep{aungwerojwit2012,worpel2018}. The subsequent detection of frequencies in the power spectrum derived from TESS lightcurves, corresponding to the 2208 s (presumed spin period) and the beat periods between the spin and orbital periods by \cite{rawat2022}, further supports the classification of V902~Mon as an IP, consistent with well-established examples of such systems \citep{warner1986}.
\par
However, the X-ray analysis of previous Swift XRT and XMM-Newton observations by \cite{aungwerojwit2012} and \cite{worpel2018} respectively, raised several questions about the nature of V902~Mon. First, spin modulation was not detected in these X-ray data, which requires explanation if V902~Mon is indeed an IP. Second, eclipses are not evident in these X-ray data, although this could be attributed to the short exposures and limited statistical quality of the data. Third, the X-ray spectrum was highly unusual and the X-ray luminosity estimated is low for an IP. We elaborate on the last point below. 
\par
The observed 0.5--8.0 keV flux of V902~Mon from Swift XRT and XMM-Newton observations of $\sim 10^{-13}$ ergs cm$^{-2}$ s$^{-1}$, which corresponds to a luminosity of $\sim$10$^{32}$ ergs s$^{-1}$ \citep{aungwerojwit2012,worpel2018}. This makes the X-ray luminosity of V902~Mon fainter than typical IPs whereas its optical to X-ray flux ratio is higher than those of typical IPs by almost two orders of magnitude \citep{aungwerojwit2012,pretorius2014}. A strong Fe K$\alpha$ line has been reported in an X-ray spectrum using an XMM-Newton observation with an equivalent width of $\sim$1.4 keV \citep{worpel2018}. This line is emitted via fluorescence when near neutral Fe atoms absorb the primary X-rays above the K edge at $\sim$7 keV. The equivalent width of the Fe K$\alpha$ emission line is a measure of the strength of the line and is defined as the ratio of the photon flux contributed by an additive model component at energy E (typically represented by a Gaussian function) to the photon flux of the continuum spectrum at the same energy. It is related to the geometry and distribution of the column density of the surrounding matter and is higher when the fluorescing matter occupies a large solid angle as seen from the primary X-ray source \citep{inoue1985,makishima1986}. In normal IPs, both the WD surface (as part of reflection -- \citealt{mukai2015}) and pre-shock accretion flow contribute to the 6.4 keV flux, and subtend large solid angles. Nevertheless, the typical equivalent width is 100--200 eV \citep{ezuka1999} and reflection models alone cannot explain the much stronger Fe K$\alpha$ fluorescence line observed in V902~Mon. 
\par
The objective of this investigation is to determine if the X-ray properties of V902~Mon are due to the edge-on viewing geometry of this eclipsing system. The observed X-rays would be a result of the scattering and reprocessing of primary X-rays occurring above the orbital plane. The eclipsing nature of V902~Mon offers a unique opportunity to study the X-ray emitting and reprocessing regions in IPs using both soft and hard X-ray observations. In this manuscript, we report results from NuSTAR and XMM-Newton observations of V902~Mon and our timing and broadband X-ray spectral analysis. The outline of the paper is as follows: in Section 2 we describe the NuSTAR, XMM and AAVSO data used for analysis. In Section 3.1, we carry out timing analysis of the energy-resolved NuSTAR lightcurves and compare them with the recent AAVSO lightcurves to search for eclipses in the hard X-ray energy-bands. In Section 3.2, we describe the broadband X-ray spectral analysis with NuSTAR and XMM observations and the results of the spectral analysis. We discuss the implications of these results, in Section 4, in understanding the accretion geometry of V902~Mon.

\section{X-ray and Optical Data and Analysis}

\begin{figure}
    \centering
    \includegraphics[scale=0.5]{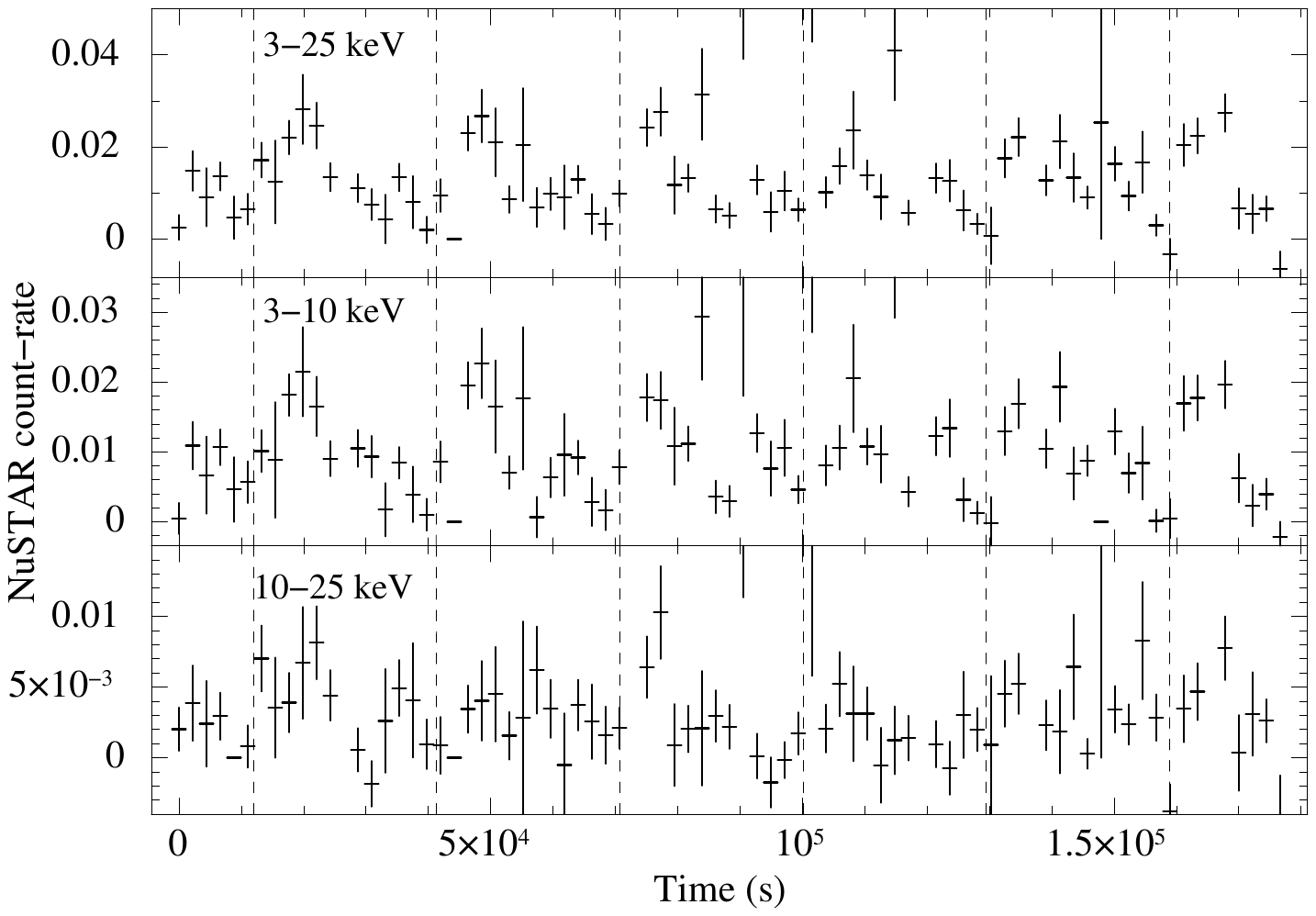}
    \caption{NuSTAR FPMA + FPMB background subtracted energy-resolved lightcurves of V902~Mon in 3--25 keV, 3--10 keV and 10--25 keV and binned by 2208 sec, the inferred spin period of the WD from optical photometry. The NuSTAR observation covered about 6 orbital cycles of the source. The dashed lines mark the expected eclipses using the ephemeris given in \cite{worpel2018}  and an orbital period of 0.34008279 d.}
    \label{nustar_lc}
\end{figure}

The {\it NuSTAR} (Nuclear Spectroscopic Telescope Array) is a hard X-ray telescope operating in 3--79 keV energy band \citep{harrison2013}. It carries two co-aligned grazing incidence Wolter I imaging telescopes that focuses onto two independent Focal Plane Modules, FPMA and FPMB. The NuSTAR observation of V902~Mon was carried out on 2023-02-13 for an effective exposure of 93 kilosec. The NuSTAR data was reduced and analyzed using NuSTAR Data Analysis Software (NuSTARDAS) v.2.1.4 package provided under {\tt HEAsoftv.6.34} and calibration files 2024-10-01. The event files were reprocessed using {\tt nupipeline} using the standard filtering procedure and the default screening criteria. The event times were corrected to the solar system barycenter using nuproducts and the FTOOL {\tt barycorr} with the DE200 solar system ephemeris. The source spectra, response matrices, ancillary matrices files and energy-resolved lightcurves were extracted in SCIENCE mode (01) from a circular region of radius 60'' centered on the source using {\tt nuproducts}. The background spectra and lightcurves were extracted from a similar circular region in a source-free region on the same chip. The lightcurves were extracted in the energy-bands 3--25 keV, 3--10 keV and 10--25 keV. Above 25 keV, background count-rates dominate the source count-rates and hence we use the NuSTAR data till 25 keV. 
\par
In addition to the NuSTAR observation, we also used an XMM-Newton observation of V902~Mon (ObsID: 0804111001), which was also used by \cite{worpel2018} for analysis. The XMM-Newton observation was carried on 2017-10-14, several years prior to the NuSTAR observations and we use only EPIC-PN data due to its high statistics compared to the MOS data. The XMM-Newton observation was carried out in Full Window Imaging mode and was analyzed using SAS v21.0.0 with the latest calibration files. We identified the intervals of flaring particle background by extracting the single event (PATTERN == 0) high energy lightcurve for EPIC-PN in 10-12 keV. The threshold for low steady background was determined to be for a RATE$\leq$0.4 counts/s for EPIC-PN. Due to high particle background flaring present during most of the observation, we could extract only about 17 ks of usable data from a total of 43 ks observation. The lightcurves and spectral files were extracted using the XMM-SAS data analysis threads\footnote{https://www.cosmos.esa.int/web/xmm-newton/sas-threads}, which filtered out the background flares and applied the latest calibration files. A circular source region was selected with a radius of 60” centered on the source and a similar size background region was selected in a source-free region on the chip. Due to lower photon statistics in the MOS observations, we use only PN data for the analysis. The event times were corrected to the solar system barycenter using {\tt barycen}. We accounted the empirical correction to the EPIC effective area based on NuSTAR observations by applying the parameter {\tt applyabsfluxcor=yes} to the XMM-SAS tool {\tt arfgen}\footnote{https://xmmweb.esac.esa.int/docs/documents/CAL-TN-0230-1-3.pdf}. The XMM/PN lightcurves were extracted in 0.3--10 keV energyband. The NuSTAR and XMM/PN source spectra were grouped to give a minimum of 30 counts in each spectral bin. The NuSTAR and XMM data analysis were carried out using {\tt HEASARCv6.34} on Sciserver \citep{taghizadeh2020}. 
\par
We downloaded 3101 individual photometric observations of V902~Mon taken by the AAVSO observers in CV-band (unfiltered data with V zeropoint) from January 2023 till March 2023\footnote{https://www.aavso.org/}, to compare with the orbital modulation and search for X-ray eclipses in the NuSTAR and XMM X-ray lightcurves. The NuSTAR observations were carried out in Feb 2023, which occur between these AAVSO observations. The timings were corrected to the solar system barycenter using {\tt astropy time} package\footnote{https://docs.astropy.org/en/stable/time/}. 
\begin{figure}
    \centering
    \includegraphics[scale=0.8]{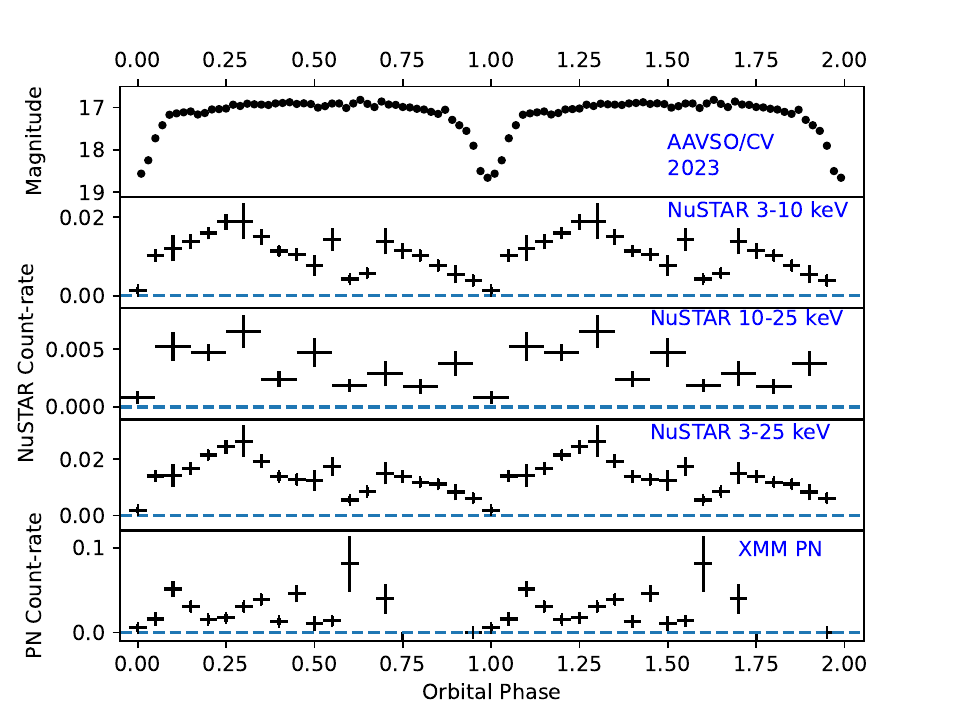}
    \caption{Orbital intensity profiles constructed using AAVSO/CV-band data for 2023 and the NuSTAR and XMM/PN background subtracted lightcurves in 3--10 keV, 10--25 keV and 0.3-10.0 keV respectively. The mid-eclipse epoch T$_{0}$, extrapolated back to Cycle 0 in 2004-12-01, is taken to be the minimum of the optical intensity, JD 2453340.5237 and the orbital period is 0.34008279 d from \cite{worpel2018}. We have used 50 phase bins for the AAVSO/CV-band orbital intensity profile and 20 phase bins for the X-ray orbital intensity profiles, except for the NuSTAR 10--25 keV orbital profile, which is constructed using 10 phase bins (due to low statistics). The blue dashed lines show zero count-rates for the NuSTAR and XMM/PN lightcurves}.
    \label{nustar_orb}
\end{figure}

\section{Results}
\subsection{Timing analysis}
Figure \ref{nustar_lc} present the NuSTAR background subtracted lightcurves of V902~Mon in 3--25 keV, 3--10 keV and 10--25 keV energy-bands and binned with 2208 sec (about 0.08 in orbital phase) which is the inferred spin of the WD from the optical photometry by \cite{worpel2018}. The NuSTAR observation covered about 6 orbital cycles and the epoch of eclipses are calculated using the ephemeris from \cite{worpel2018} and are marked by dashed lines in Figure \ref{nustar_lc}. We do not see any strong variability in the NuSTAR lightcurves, except for the orbital modulation which is mentioned in detail in subsequent paragraphs.
\par
We carried out a search for pulsations using Epoch Folding (EF) and $\chi^{2}$ maximization techniques \citep{leahy1987} and $Z^{2}$ pulsation search \citep{buccheri1983}, on the cleaned NuSTAR unbinned event files, but did not find any strong peaks in the EF or $Z^{2}$ periodogram except at period of the gaps in the NuSTAR data ($\sim$97 minutes) and its harmonics. This is consistent with the non-detection of pulsations from the previous observations with {\it Swift} XRT and XMM-Newton by \cite{worpel2018}. The pulse fraction is defined as:\\
\begin{equation}
    PF = \frac{F_{max}-F_{min}}{F_{max}+F_{min}} \nonumber
\end{equation}
where $F_{\rm{max}}$ and $F_{\rm{min}}$ are the maximum and minimum count-rates in the pulse profile respectively. The upper limit on the pulse fraction estimated by folding the NuSTAR 3--25 keV lightcurve on the optical spin period of 2208 s using 16 phase bins, is 32\%.
\par
The AAVSO/CV-band lightcurves, the NuSTAR lightcurves in 3--10 keV, 10--25 keV, 3--25 keV energy bands and the XMM/PN lightcurves in 0.3-10 keV energy band were folded with the orbital period of 0.34008279 d and plotted in Figure \ref{nustar_orb}. The mid-eclipse epoch T$_{0}$ is chosen as the minimum of the optical intensity, as shown in the top panel of Figure \ref{nustar_orb}. This corresponds to a updated mid-eclipse epoch of JD 2453340.5237\footnote{This is the nominal mid-eclipse epoch extrapolated back to Cycle 0 on 2004-12-01 using a linear ephemeris with P=0.34008279 d; given the period derivative in \cite{rawat2022}, this diverges from the mid-eclipse times reported by \cite{worpel2018} by about 24 minutes.}, which is consistent within the error bars and the orbital period derivative for this system \citep{worpel2018, rawat2022}. We see that a phase bin corresponding to the lowest count-rate for the NuSTAR orbital intensity profiles coincides with the mid-eclipse phase from the AAVSO/CV-band orbital profile. Since the time of mid-eclipse in the optical is when the WD is exactly behind the secondary star, this indicates that the X-ray emission region is also centered on the WD, as expected.  The count-rates and error-bars of the phase bin of the NuSTAR energy-resolved orbital intensity profiles coincident with the optical mid-eclipse epoch are 0.0013$\pm$0.0009 counts/sec for 3--10 keV, 0.0008$\pm$0.0006 counts/sec for 10--25 keV, and 0.002$\pm$0.001 counts/sec for 3--25 keV. It is difficult to distinguish if these low count-rates indicate a total eclipse, or if there is a low-level uneclipsed X-ray emission. In either case, this is the first confirmation of an eclipse in X-rays for V902~Mon.  
\par
The XMM observation suffered from high background flaring and the filtered PN event files only has 17 ks of usable data which is less than one complete orbital cycle. The cleaned PN lightcurve has a gap in the data between orbital phase 0.7 to 0.95 using our updated ephemeris (see the bottom panel of Figure \ref{nustar_orb}). However, this updated ephemeris cannot reliably be extrapolated back to 2017 without a full and accurate knowledge of the orbital period derivative (see Figure 6 of \citealt{rawat2022}).  If we use the ephemeris from \cite{worpel2018} instead, the gap in the lightcurve is between orbital phase 0.75 and 1.0. Therefore, we conclude that the background flaring in the XMM/PN observation precludes us from reaching a reliable conclusion regarding the presence of otherwise of an X-ray eclipse in the XMM observation. 
\par
The eclipse in the optical AAVSO/CV-band orbital folded lightcurve is deep, with about a factor of 6 difference in brightness in and out of the eclipse and an eclipse half-width of about orbital phase of 0.1. In comparison, the change in the count-rates in and out-of-eclipse for NuSTAR lightcurves is about a factor of 10--20 at a minimum, taking the apparent mid-eclipse residual X-ray flux at face value. The X-ray eclipse is narrower compared to optical profile, with only one phase bin corresponding to the minimum X-ray count-rates. This is understandable if the X-ray emitting region is compact as expected, with a size similar to the WD, whereas the optical emitting region (the accretion disk) occupies a substantial fraction of the primary Roche lobe. In addition to the eclipse at orbital phase 0, there appears to be an out-of-eclipse orbital modulation of X-ray intensities with a peak at phase 0.3 and secondary dip in the intensity at phase 0.6. Both the X-ray eclipse and the broader orbital modulation in the X-rays are interpreted in detail in Section 4. 

\subsection{X-ray spectral Fitting}
\begin{figure}
    \centering
    \includegraphics[scale=0.5]{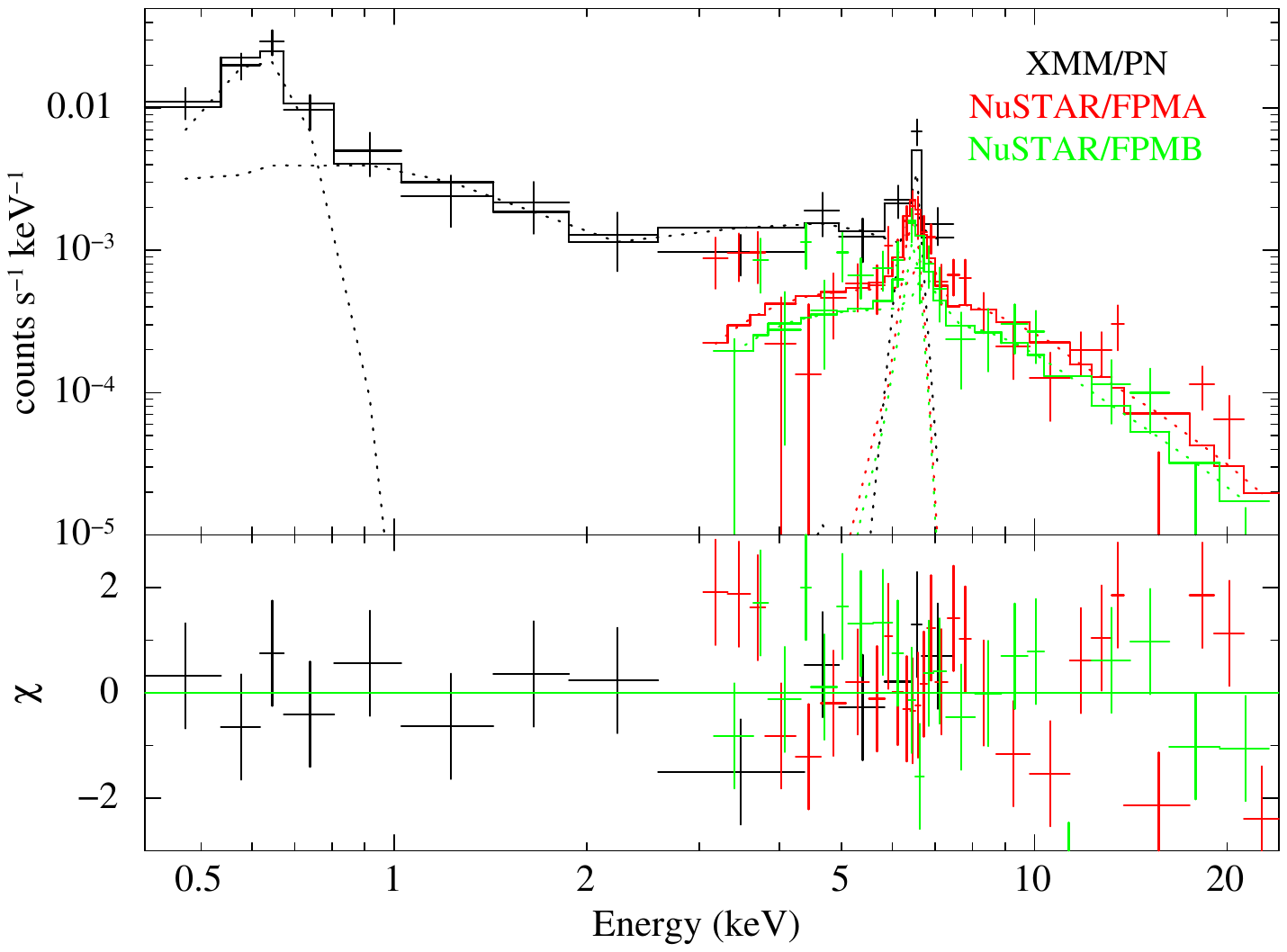}
    \caption{Broadband X-ray spectral fits using XMM and NuSTAR observations (top panel) and residuals to the spectral fits (bottom panel). The spectral model for the fits is described in Section 3.2. The spectra are rebinned for clarity}
    \label{nustar_spec}
\end{figure}

The X-ray broadband spectral fitting was carried out in the energy range 0.3--10 keV for XMM/PN and 3--25 keV for NuSTAR FPMA and FPMB. \cite{worpel2018} modelled the XMM MOS 1, MOS 2 and PN spectra with a partially covered two {\tt mekal} models, with one temperature estimated at 194 eV and other at 15.4 keV, and a Gaussian line for the strong 6.4 Fe K$\alpha$ fluorescence line. The lower plasma temperature accounted for the soft excess seen in the XMM/PN spectrum. Since we expect the X-ray emission from IPs to be a resultant emission from plasma over a continuous temperature distribution, from the shock temperature to the WD photospheric temperature, we model the broadband spectra using a cooling flow model {\tt mkcflow}, modified by a fully covering photo-electric absorption model {\tt Tbabs}, a partial covering absorption model {\tt Tbpcf} and a Gaussian line at 6.4 keV \citep{mukai2003a}. To account for the soft excess seen in the XMM/PN spectrum at lower energies $\leq$0.7 keV, we model it with an additional Gaussian for a blend of emission lines including O VII emission line at 0.556 keV, which is modified by only the fully covering photo-electric absorption model {\tt Tbabs}. A constant was added for the XMM/PN and NuSTAR FPMA + FPMB spectral fit to account for cross-normalization difference between the different instruments. This constant of cross-normalization would also account for the small changes in the X-ray fluxes in the XMM and NuSTAR observations as they were carried out at different times. All spectral fits were carried out using {\tt XSPEC v12.14.1} \citep{arnaud1996}.
The full spectral model in {\tt XSPEC} is: {\tt constant*Tbabs*(Tbpcf*(mkcflow+gauss(6.4 keV)))+gauss(0.55 keV))}. Table 1 provides the spectral parameters of the best fit to the XMM/PN and NuSTAR spectra. The error on the spectral parameters are calculated at 90\% confidence limits. The X-ray fluxes and their errors were estimated using the {\tt cflux} model in {\tt XSPEC}.

\begin{table*}
\centering
\caption{Best fitting parameter values for the XMM + NuSTAR spectra of V902~Mon using the model defined in Section 3.2. The errors on the parameters are estimated using 90\% confidence limits}
\begin{tabular}{c c c}
\hline
$C_{\rm{FPMA}}$ & & 1.0 (fixed) \\
$C_{\rm{FPMB}}$ & & 0.8$\pm$0.2 \\
$C_{\rm{PN}}$ & & 0.7$\pm$0.2 \\
\nh$_{1}$ & 10$^{22}$cm$^{-2}$ & 0.04 (fixed) \\
\nh$_{2}$ & 10$^{22}$cm$^{-2}$ & 14$^{+7}_{-5}$ \\
CvrFrac & & 0.91$^{+0.04}_{-0.06}$ \\
$kT_{\rm {max}}$ & keV & 16$^{+12}_{-6}$ \\
Normalization & M$_\odot$/yr & (4.1$^{+4}_{-2}$) $\times 10^{-11}$ \\
O VII & keV & 0.56$\pm$0.03 \\
Norm (O VII) & photons/cm$^{2}$/s & (2.1$\pm$0.1) $\times 10^{-3}$ \\
Fe K$\alpha$ & keV & 6.38$\pm$0.07 \\
Norm (Fe K$\alpha$) & photons/cm$^{2}$/s & (4$\pm$1) $\times 10^{-6}$ \\
Eqw (Fe K$\alpha$) & keV & 0.67$\pm$0.05 keV \\
Eqw (O VII) & keV & 1.49$\pm$0.07 keV \\
$\chi^{2}$ & & 96 for 88 d.o.f \\
Unabsorbed Flux & 0.3-8.0 keV & (1.7$\pm$0.1) $\times 10^{-13}$ \\
Unabsorbed Flux & 3-25 keV & (2.9$\pm$0.3) $\times 10^{-13}$ \\
\hline
\end{tabular}
\end{table*}

\par
The absorption column density with a covering fraction of 0.9, is of the order of $10^{23}$ cm$^{-2}$, which suggests a strong local absorption around the source for a modest absorbed X-ray flux of $10^{-13}$ ergs/s/cm$^{-2}$. At the estimated distance of V902~Mon, the 3D extinction map of \cite{doroshenko2024}\footnote{http://astro.uni-tuebingen.de/nh3d/nhtool} based on optical-IR reddening of stars with Gaia distances shows E(B-V): 0.63$\pm$0.1 mag, which, using the canonical relationship, implies an interstellar \nh\ of $\sim 4 \times 10^{21}$ cm$^{-2}$. 
Such a high \nh\ has an exponential cut-off around 0.7 keV, yet we detect softer X-rays from V902~Mon. This implies that either V902~Mon has an extraordinarily luminous soft X-ray component that can shine through an high interstellar \nh\ of $\sim 4 \times 10^{21}$, or that the interstellar \nh\ is overestimated when using the Gaia-based 3D extinction maps. We argue that the latter is a possibility, because a wide range of gas-to-dust ratio is observed in individual clouds (see, e.g., \citealt{chen2015}). Therefore, we assume that the ISM in this direction has a high dust content (leading to high optical/IR extinction) but relatively little gas content (leading to less attenuation of soft X-rays). We present our X-ray spectral fit with a much lower interstellar \nh\ of $\sim 4 \times 10^{20}$, so as not to require an extraordinarily luminous soft component (this issue is common to model used here as well the model in \citealt{worpel2018}), and leave the issue of true interstellar \nh\ versus true soft luminosity of V902~Mon to future studies.
\par
A strong Fe K$\alpha$ fluorescence line is present in the XMM/PN and NuSTAR spectra, with an equivalent width of about 0.7 keV.  In addition, we modeled the soft excess seen in the PN spectrum using a Gaussian emission line, which models the blend of emission lines at 0.56 keV including OVII. The equivalent width of this Gaussian line is about 1.5 keV. 

\section{Discussions}
\begin{figure}
    \centering
    \includegraphics[scale=0.8]{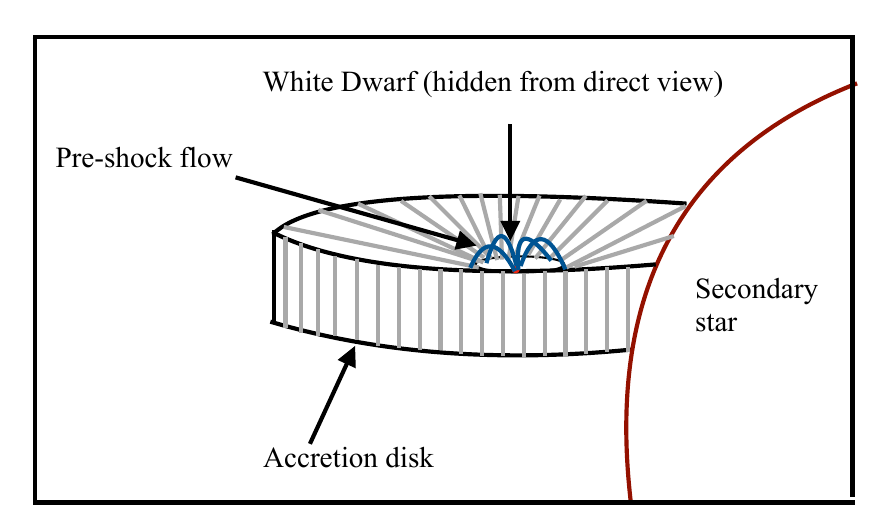}
    \caption{Schematic diagram of V902~Mon. The WD is hidden from the direct view at all orbital phases by the thick accretion disk. The X-ray eclipse is caused by the occultation of the X-ray emitting region, like the pre-shock region located few $R_{\rm{wd}}$ above the orbital plane, by the secondary star.}
    \label{diagram}
\end{figure}
As shown in Figure \ref{nustar_orb}, the NuSTAR energy-resolved light curves confirm the presence of an X-ray eclipse in V902~Mon. Through broadband X-ray spectral analysis combining XMM-Newton and NuSTAR observations, we also confirm a strong Fe K$\alpha$ line with a high equivalent width, consistent with previous findings by \cite{worpel2018} based on XMM-Newton data alone. While direct evidence of V902~Mon's IP nature via the detection of X-ray spin modulation remains elusive, our results indirectly support the IP classification, as discussed below. Moreover, we propose that V902~Mon is likely a typical IP observed at high inclination, surrounded by dense absorbing material, likely originating from its accretion disk.
\par
The unabsorbed X-ray flux of 1.7$\times$10$^{-13}$ ergs\,cm$^{-2}$\,s$^{-1}$ in 0.3-8.0 keV and 2.9$\times$10$^{-13}$ ergs\,cm$^{-2}$\,s$^{-1}$ in 3--25 keV estimates the X-ray luminosities of 1.8$\times$10$^{32}$ ergs\,s$^{-1}$ and 3.1$\times$10$^{32}$ ergs\,s$^{-1}$ respectively, for a distance of 3 kpc. This is higher than the X-ray luminosity of nearby quiescent dwarf novae \citep{byckling2010}, although confirmed and likely non-magnetic CVs do exist with comparable or higher X-ray luminosities ({\it e.g}. BV Cen at $\sim 3 \times 10^{32}$ ergs s$^{-1}$ -- \citealt{wada2017}). Furthermore, the high equivalent width of the Fe K$\alpha$ fluorescence line likely implies an hidden X-ray component whose luminosity is perhaps an order of magnitude higher than that of the observed component. This is because the observed absorbing column of the partial-covering absorber (\nh\ $\sim 10^{23}$ cm$^{-2}$) is not high enough to produce the strong Fe K$\alpha$ fluorescence line (equivalent width of $\sim$0.67 keV) that we observe (\citealt{ezuka1999}; see also \citealt{inoue1985, ishida1991}).  A likely explanation is that the primary X-ray continuum that is responsible for this fluorescent line is hidden from our view.  This is similar to the case with High-Mass X-ray binaries during the X-ray eclipse when we observe X-rays reprocessed by the surrounding medium situated at larger distance from the companion star whereas the direct X-rays are blocked by the companion star \citep{aftab2019}. Similarly, the lower X-ray flux and high equivalent width of the Fe K$\alpha$ line observed with Suzaku in the symbiotic system, CH Cyg, was interpreted by \cite{mukai2007} as due to the primary emission site being hidden from our view by the body of the accretion disk. 
\par
We therefore consider the following scenario to explain the X-ray characteristics of V902~Mon, as shown in the schematic diagram in Figure \ref{diagram}. The WD is always hidden from our view by the body of the accretion disk, with \nh\ about 10$^{24}$ cm$^{-2}$ or higher, well into the Compton-thick regime. Therefore the X-ray photons are absorbed and scattered away from our line of sight by this Compton-thick absorbing medium and we do not see direct X-ray emission. All the X-rays we do observe have been reprocessed or scattered into our line of sight, mostly likely in a region above the orbital plane. Therefore, the intrinsic X-ray luminosity of V902~Mon is perhaps an order of magnitude higher than the luminosity of the observed component. This scenario suggests that the X-ray luminosity of V902~Mon in about  an order of magnitude higher than the observed luminosity of about 10$^{32}$ ergs\,s$^{-1}$, which makes this system as luminous as typical IPs and well above the typical X-ray luminosities of confirmed non-magnetic CVs \citep{mukai2023}. As the pre-shock accretion flow in IPs typically occur above the orbital plane, this is a possible location for the X-ray reprocessing region.
\par
This scenario could also account for the high column density observed in the X-ray spectrum and the broad, out-of-eclipse orbital modulation shown in Figure \ref{nustar_orb}. The accretion disk is not a solid body with a sharp boundary, for which there is a clear and unambiguous definition of height, above which no disk materials exist. The height of a disk is a scale height; there is material above the scale height of a disk, albeit at reduced densities. We propose that our line of sight to the X-ray reprocessing site passes through the outer regions (hereafter referred to as the ``atmosphere") of the accretion disk. Additionally, the height of the outer accretion disk is not uniform but exhibits azimuthal variation. This picture is similar to the Accretion Disk Corona (ADC) Low-Mass X-ray Binaries (LMXBs), which are observed at very high inclinations (edge-on view). In these systems, the line of sight to the compact object is obscured by the accretion disk itself. ADC sources are characterized by lower apparent X-ray luminosities compared to typical LMXBs. The X-rays are scattered from the corona which lie above the orbital plane, but obscuration by the outer disk rim produces an orbital modulation of the X-ray intensity \citep{hellier1989}. 
\par
The X-ray spin modulation of IPs is related to variable complex absorption in the pre-shock flow (see \citealt{islam2021} and references therein). This mechanism is unlikely to be effective in creating X-ray spin modulation when all the X-rays we observe are scattered in the pre-shock flow itself. The lack of detection of pulsations in the NuSTAR observations would likely be due to either the scattering of the X-rays in the pre-shock flow and/or the low signal-to-noise ratio of the X-ray data and is not an evidence against the IP interpretation of this source.
\par
The orbital geometry inferred from optical observations by \cite{aungwerojwit2012} indicates a high inclination for the system, suggesting that any X-ray emitting or reprocessing region would be eclipsed by the secondary star. This is confirmed by the detection of the X-ray eclipse. The orbital intensity profiles from the NuSTAR lightcurves suggest that the X-ray eclipse is total and occurs in phase with the optical eclipse, though statistical limitations prevent definitive confirmation. Additionally, the exact width of the X-ray eclipse remains poorly constrained. However, it is evident that X-ray ingress begins after the optical ingress, implying that the X-ray reprocessing region does not extend to the outer edges of the accretion disk. Precise determination of the orbital geometry will require higher signal-to-noise X-ray data across both soft and hard X-ray bands.
\par
The equivalent width of the Gaussian line, used to model the soft excess and interpreted as a blend of emission lines around 0.56 keV (including OVII), is approximately 1.5 keV. This value is notably high, even within the scattering/reprocessing framework. We propose that some or all of the OVII line photons may originate at a greater height above the orbital plane than the scattered continuum, thereby escaping the absorption by the accretion disk atmosphere that the X-ray continuum goes through.
\par
Finally, we consider V902~Mon in the context of nine confirmed IPs that exhibit varying degrees of eclipsing behavior\footnote{This excludes V597~Pup, which is a deeply eclipsing CV that was proposed to be an IP by \citet{warner2009}, but is not confirmed.}. There are three IPs, FO~Aqr \citep{hellier1990}, BG~CMi \citep{patterson1993} and TV~Col \citep{hellier1991} that show grazing eclipses in optical of the outer accretion disk. The inclination angle is not high enough in these systems for the WD to be eclipsed, hence we do not observe an X-ray eclipse in them.  EX~Hya exhibits partial eclipses in both optical and X-rays \citep{gilliland1982, beuermann1985}. The partial eclipses require a higher inclination than for FO~Aqr, BG~CMi or TV~Col. There are three IPs with total X-ray eclipses. Of these, IGR~J17014$-$4306 and CXOGBS J174954.5--294335 are deeply eclipsing both in the optical and in X-rays \citep{shara2017,bernardini2017,johnson2017}. The other, XY~Ari is deeply eclipsing in X-rays but is not visible in the optical band because it is hidden behind a molecular cloud MBM12 \citep{patterson1990,hellier1997}. In these systems, the inclination is clearly high enough for the WD to be eclipsed, but not so high as for our line of sight to be blocked by the body of the accretion disk outside of the eclipse. Finally, DQ~Her is the IP that most closely resembles our proposed interpretation of V902~Mon. DQ~Her is deeply eclipsing in the optical \citep{walker1954} and underluminous in X-rays, with a shallow partial X-ray eclipse \citep{mukai2003b}. The differences between V902~Mon and DQ~Her can be explained as a consequence of the much shorter spin period (71 s) of the latter: the pre-shock accretion flow is much more compact, making it far easier for it to be hidden by the body of the accretion disk, whereas for V902~Mon the pre-shock region is likely extended few $R_{\rm{wd}}$ above the orbital plane.

\section{Summary}
We carried out timing and spectral analysis of the eclipsing IP V902~Mon using X-ray observations with NuSTAR and XMM-Newton and optical observations with the AAVSO/CV-band lightcurves. We find a clear evidence of an X-ray eclipse in the NuSTAR energy-resolved lightcurve, which establishes this source as an eclipsing CV in X-rays. We do not detect any pulsations in the NuSTAR lightcurves. Using XMM and NuSTAR observations, we carried out broadband X-ray spectral modeling and confirm the presence of a strong Fe K$\alpha$ line at 6.4 keV and an absorption column density of 10$^{23}$ cm$^{-2}$ along the line of sight towards the X-ray emitting region. The 0.7 keV equivalent width of the Fe fluorescence line suggests the presence of a higher luminosity component, hidden from our view by the body of the accretion disk. We propose an ADC-source like picture, in which V902~Mon is an IP whose primary X-ray component is always hidden from our view, and we observed X-ray scattered/reprocessed in the pre-shock accretion flow a few $R_{\rm{wd}}$ above the orbital plane.

\medskip\noindent{\bf Acknowledgments:}\\
We thank the anonymous referee for constructive comments which helped improve the manuscript.
This work was supported by NASA grant 80NSSC21K0022 and RSA No. 1698067. The scientific results reported here are based on observations made by the NuSTAR X-ray observatory, and we thank the NuSTAR Operations,
Software, and Calibration teams for scheduling and the execution of these observations. This research has made use of NuSTAR Data Analysis Software (NuSTARDAS) jointly developed by the ASI Science Science Data Center (ASDC, Italy) and the California Institute of Technology. This research has made use of data and/or software provided by the High Energy Astrophysics Science Archive Research Center (HEASARC), which is a service of the Astrophysics Science Division at NASA/GSFC. This research makes use of the SciServer science platform (www.sciserver.org). 
SciServer is a collaborative research environment for large-scale data-driven science. It is being developed at, and administered by, the Institute for Data Intensive Engineering and Science at Johns Hopkins University. SciServer is funded by the National Science Foundation through the Data Infrastructure Building Blocks (DIBBs) program and others, as well as by the Alfred P. Sloan Foundation and the Gordon and Betty Moore Foundation. We have extracted archival data from the XMM-Newton Science Archive and the data analysis was performed with the XMM-Newton SAS. We acknowledge with thanks the variable star observations from the AAVSO International Database contributed by observers worldwide and used in this research.

\bibliography{bibtex}

\begin{thebibliography}{}
\expandafter\ifx\csname natexlab\endcsname\relax\def\natexlab#1{#1}\fi
\providecommand{\url}[1]{\href{#1}{#1}}
\providecommand{\dodoi}[1]{doi:~\href{http://doi.org/#1}{\nolinkurl{#1}}}
\providecommand{\doeprint}[1]{\href{http://ascl.net/#1}{\nolinkurl{http://ascl.net/#1}}}
\providecommand{\doarXiv}[1]{\href{https://arxiv.org/abs/#1}{\nolinkurl{https://arxiv.org/abs/#1}}}

\bibitem[{{Aftab} {et~al.}(2019){Aftab}, {Paul}, \& {Kretschmar}}]{aftab2019}
{Aftab}, N., {Paul}, B., \& {Kretschmar}, P. 2019, {X-Ray Reprocessing: Through
  the Eclipse Spectra of High-mass X-Ray Binaries with XMM-Newton}, \apjs, 243,
  29, \dodoi{10.3847/1538-4365/ab2a77}

\bibitem[{{Arnaud}(1996)}]{arnaud1996}
{Arnaud}, K.~A. 1996, in Astronomical Society of the Pacific Conference Series,
  Vol. 101, Astronomical Data Analysis Software and Systems V, ed. G.~H.
  {Jacoby} \& J.~{Barnes}, 17

\bibitem[{{Aungwerojwit} {et~al.}(2012){Aungwerojwit}, {G{\"a}nsicke},
  {Wheatley}, {Pyrzas}, {Staels}, {Krajci}, \&
  {Rodr{\'\i}guez-Gil}}]{aungwerojwit2012}
{Aungwerojwit}, A., {G{\"a}nsicke}, B.~T., {Wheatley}, P.~J., {et~al.} 2012,
  {IPHAS J062746.41+014811.3: A Deeply Eclipsing Intermediate Polar}, \apj,
  758, 79, \dodoi{10.1088/0004-637X/758/2/79}

\bibitem[{{Bailer-Jones} {et~al.}(2021){Bailer-Jones}, {Rybizki}, {Fouesneau},
  {Demleitner}, \& {Andrae}}]{bailerjones2021}
{Bailer-Jones}, C.~A.~L., {Rybizki}, J., {Fouesneau}, M., {Demleitner}, M., \&
  {Andrae}, R. 2021, {Estimating Distances from Parallaxes. V. Geometric and
  Photogeometric Distances to 1.47 Billion Stars in Gaia Early Data Release 3},
  \aj, 161, 147, \dodoi{10.3847/1538-3881/abd806}

\bibitem[{{Bernardini} {et~al.}(2017){Bernardini}, {de Martino}, {Mukai},
  {Russell}, {Falanga}, {Masetti}, {Ferrigno}, \& {Israel}}]{bernardini2017}
{Bernardini}, F., {de Martino}, D., {Mukai}, K., {et~al.} 2017, {Broad-band
  characteristics of seven new hard X-ray selected cataclysmic variables},
  \mnras, 470, 4815, \dodoi{10.1093/mnras/stx1494}

\bibitem[{{Beuermann} \& {Osborne}(1985)}]{beuermann1985}
{Beuermann}, K., \& {Osborne}, J. 1985, {Hard X-Ray Observations of the
  Eclipsing Binary Ex-Hydrae}, \ssr, 40, 117, \dodoi{10.1007/BF00212873}

\bibitem[{{Buccheri} {et~al.}(1983){Buccheri}, {Bennett}, {Bignami}, {Bloemen},
  {Boriakoff}, {Caraveo}, {Hermsen}, {Kanbach}, {Manchester}, {Masnou},
  {Mayer-Hasselwander}, {{\"O}zel}, {Paul}, {Sacco}, {Scarsi}, \&
  {Strong}}]{buccheri1983}
{Buccheri}, R., {Bennett}, K., {Bignami}, G.~F., {et~al.} 1983, {Search for
  pulsed {\ensuremath{\gamma}}-ray emission from radio pulsars in the COS-B
  data.}, \aap, 128, 245

\bibitem[{{Byckling} {et~al.}(2010){Byckling}, {Mukai}, {Thorstensen}, \&
  {Osborne}}]{byckling2010}
{Byckling}, K., {Mukai}, K., {Thorstensen}, J.~R., \& {Osborne}, J.~P. 2010,
  {Deriving an X-ray luminosity function of dwarf novae based on parallax
  measurements}, \mnras, 408, 2298, \dodoi{10.1111/j.1365-2966.2010.17276.x}

\bibitem[{{Chen} {et~al.}(2015){Chen}, {Liu}, {Yuan}, {Huang}, \&
  {Xiang}}]{chen2015}
{Chen}, B.~Q., {Liu}, X.~W., {Yuan}, H.~B., {Huang}, Y., \& {Xiang}, M.~S.
  2015, {Dust-to-gas ratio, X$_{CO}$ factor and CO-dark gas in the Galactic
  anticentre: an observational study}, \mnras, 448, 2187,
  \dodoi{10.1093/mnras/stv103}

\bibitem[{{Doroshenko}(2024)}]{doroshenko2024}
{Doroshenko}, V. 2024, {3D-$N_{\rm H}$-tool}, arXiv e-prints, arXiv:2403.03127,
  \dodoi{10.48550/arXiv.2403.03127}

\bibitem[{{Ezuka} \& {Ishida}(1999)}]{ezuka1999}
{Ezuka}, H., \& {Ishida}, M. 1999, {Iron Line Diagnostics of the Postshock Hot
  Plasma in MagneticCataclysmic Variables Observed with ASCA}, \apjs, 120, 277,
  \dodoi{10.1086/313181}

\bibitem[{{Gilliland}(1982)}]{gilliland1982}
{Gilliland}, R.~L. 1982, {EX Hya : physical parameters derived from
  simultaneous spectroscopy and photometry.}, \apj, 258, 576,
  \dodoi{10.1086/160109}

\bibitem[{{Harrison} {et~al.}(2013){Harrison}, {Craig}, {Christensen},
  {Hailey}, {Zhang}, {Boggs}, {Stern}, {Cook}, {Forster}, {Giommi},
  {Grefenstette}, {Kim}, {Kitaguchi}, {Koglin}, {Madsen}, {Mao}, {Miyasaka},
  {Mori}, {Perri}, {Pivovaroff}, {Puccetti}, {Rana}, {Westergaard}, {Willis},
  {Zoglauer}, {An}, {Bachetti}, {Barri{\`e}re}, {Bellm}, {Bhalerao},
  {Brejnholt}, {Fuerst}, {Liebe}, {Markwardt}, {Nynka}, {Vogel}, {Walton},
  {Wik}, {Alexander}, {Cominsky}, {Hornschemeier}, {Hornstrup}, {Kaspi},
  {Madejski}, {Matt}, {Molendi}, {Smith}, {Tomsick}, {Ajello}, {Ballantyne},
  {Balokovi{\'c}}, {Barret}, {Bauer}, {Blandford}, {Brandt}, {Brenneman},
  {Chiang}, {Chakrabarty}, {Chenevez}, {Comastri}, {Dufour}, {Elvis}, {Fabian},
  {Farrah}, {Fryer}, {Gotthelf}, {Grindlay}, {Helfand}, {Krivonos}, {Meier},
  {Miller}, {Natalucci}, {Ogle}, {Ofek}, {Ptak}, {Reynolds}, {Rigby},
  {Tagliaferri}, {Thorsett}, {Treister}, \& {Urry}}]{harrison2013}
{Harrison}, F.~A., {Craig}, W.~W., {Christensen}, F.~E., {et~al.} 2013, {The
  Nuclear Spectroscopic Telescope Array (NuSTAR) High-energy X-Ray Mission},
  \apj, 770, 103, \dodoi{10.1088/0004-637X/770/2/103}

\bibitem[{{Hellier}(1996)}]{hellier1996}
{Hellier}, C. 1996, in Astrophysics and Space Science Library, Vol. 208, IAU
  Colloq. 158: Cataclysmic Variables and Related Objects, ed. A.~{Evans} \&
  J.~H. {Wood}, 143, \dodoi{10.1007/978-94-009-0325-8_44}

\bibitem[{{Hellier}(1997)}]{hellier1997}
{Hellier}, C. 1997, {The size of the accretion region in intermediate polars:
  eclipses of XY ARIETIS observed with RXTE}, \mnras, 291, 71,
  \dodoi{10.1093/mnras/291.1.71}

\bibitem[{{Hellier} \& {Mason}(1989)}]{hellier1989}
{Hellier}, C., \& {Mason}, K.~O. 1989, {EXOSAT observations of X 1822-371 :
  modelling of the accretion disc rim.}, \mnras, 239, 715,
  \dodoi{10.1093/mnras/239.3.715}

\bibitem[{{Hellier} {et~al.}(1990){Hellier}, {Mason}, \&
  {Cropper}}]{hellier1990}
{Hellier}, C., {Mason}, K.~O., \& {Cropper}, M. 1990, {Spectroscopy of the
  intermediate polar FO Aquarii.}, \mnras, 242, 250,
  \dodoi{10.1093/mnras/242.2.250}

\bibitem[{{Hellier} {et~al.}(1991){Hellier}, {Mason}, \&
  {Mittaz}}]{hellier1991}
{Hellier}, C., {Mason}, K.~O., \& {Mittaz}, J.~P.~D. 1991, {An eclipse in the
  quadruple-period intermediate polar TV Columbae.}, \mnras, 248, 5P,
  \dodoi{10.1093/mnras/248.1.5P}

\bibitem[{{Inoue}(1985)}]{inoue1985}
{Inoue}, H. 1985, {TENMA Observations of Bright Binary X-Ray Sources}, \ssr,
  40, 317, \dodoi{10.1007/BF00212905}

\bibitem[{{Ishida}(1991)}]{ishida1991}
{Ishida}, M. 1991, PhD thesis, -

\bibitem[{{Islam} \& {Mukai}(2021)}]{islam2021}
{Islam}, N., \& {Mukai}, K. 2021, {The Role of Complex Ionized Absorbers in the
  Soft X-Ray Spectra of Intermediate Polars}, \apj, 919, 90,
  \dodoi{10.3847/1538-4357/ac134e}

\bibitem[{{Johnson} {et~al.}(2017){Johnson}, {Torres}, {Hynes}, {Jonker},
  {Heinke}, {Maccarone}, {Britt}, {Steeghs}, {Wevers}, \& {Wu}}]{johnson2017}
{Johnson}, C.~B., {Torres}, M.~A.~P., {Hynes}, R.~I., {et~al.} 2017, {CXOGBS
  J174954.5-294335: a new deeply eclipsing intermediate polar}, \mnras, 466,
  129, \dodoi{10.1093/mnras/stw3063}

\bibitem[{{Leahy}(1987)}]{leahy1987}
{Leahy}, D.~A. 1987, {Searches for pulsed emission - Improved determination of
  period and amplitude from epoch folding for sinusoidal signals}, \aap, 180,
  275

\bibitem[{{Makishima}(1986)}]{makishima1986}
{Makishima}, K. 1986, in The Physics of Accretion onto Compact Objects, ed.
  K.~O. {Mason}, M.~G. {Watson}, \& N.~E. {White}, Vol. 266, 249,
  \dodoi{10.1007/3-540-17195-9_14}

\bibitem[{{Mukai}(2017)}]{mukai2017}
{Mukai}, K. 2017, {X-Ray Emissions from Accreting White Dwarfs: A Review},
  \pasp, 129, 062001, \dodoi{10.1088/1538-3873/aa6736}

\bibitem[{{Mukai} {et~al.}(2007){Mukai}, {Ishida}, {Kilbourne}, {Mori},
  {Terada}, {Chan}, \& {Soong}}]{mukai2007}
{Mukai}, K., {Ishida}, M., {Kilbourne}, C., {et~al.} 2007, {An Apparent Hard
  X-Ray Decline of CH Cygni}, \pasj, 59, 177, \dodoi{10.1093/pasj/59.sp1.S177}

\bibitem[{{Mukai} {et~al.}(2003{\natexlab{a}}){Mukai}, {Kinkhabwala},
  {Peterson}, {Kahn}, \& {Paerels}}]{mukai2003a}
{Mukai}, K., {Kinkhabwala}, A., {Peterson}, J.~R., {Kahn}, S.~M., \& {Paerels},
  F. 2003{\natexlab{a}}, {Two Types of X-Ray Spectra in Cataclysmic Variables},
  \apjl, 586, L77, \dodoi{10.1086/374583}

\bibitem[{{Mukai} \& {Pretorius}(2023)}]{mukai2023}
{Mukai}, K., \& {Pretorius}, M.~L. 2023, {The orbital period versus absolute
  magnitude relationship of intermediate polars: implications for low states
  and outbursts}, \mnras, 523, 3192, \dodoi{10.1093/mnras/stad1603}

\bibitem[{{Mukai} {et~al.}(2015){Mukai}, {Rana}, {Bernardini}, \& {de
  Martino}}]{mukai2015}
{Mukai}, K., {Rana}, V., {Bernardini}, F., \& {de Martino}, D. 2015,
  {Unambiguous Detection of Reflection in Magnetic Cataclysmic Variables: Joint
  NuSTAR-XMM-Newton Observations of Three Intermediate Polars}, \apjl, 807,
  L30, \dodoi{10.1088/2041-8205/807/2/L30}

\bibitem[{{Mukai} {et~al.}(2003{\natexlab{b}}){Mukai}, {Still}, \&
  {Ringwald}}]{mukai2003b}
{Mukai}, K., {Still}, M., \& {Ringwald}, F.~A. 2003{\natexlab{b}}, {The Origin
  of Soft X-Rays in DQ Herculis}, \apj, 594, 428, \dodoi{10.1086/376752}

\bibitem[{{Patterson}(1994)}]{patterson1994}
{Patterson}, J. 1994, {The DQ Herculis Stars}, \pasp, 106, 209,
  \dodoi{10.1086/133375}

\bibitem[{{Patterson} \& {Halpern}(1990)}]{patterson1990}
{Patterson}, J., \& {Halpern}, J.~P. 1990, {On the Nature of the X-Ray Pulsar
  near Lynds 1457}, \apj, 361, 173, \dodoi{10.1086/169180}

\bibitem[{{Patterson} \& {Thomas}(1993)}]{patterson1993}
{Patterson}, J., \& {Thomas}, G. 1993, {Rapid Oscillations in Cataclysmic
  Variables. IX. BG Canis Minoris (=3A 0729+103)}, \pasp, 105, 59,
  \dodoi{10.1086/133127}

\bibitem[{{Pretorius} \& {Mukai}(2014)}]{pretorius2014}
{Pretorius}, M.~L., \& {Mukai}, K. 2014, {Constraints on the space density of
  intermediate polars from the Swift-BAT survey}, \mnras, 442, 2580,
  \dodoi{10.1093/mnras/stu990}

\bibitem[{{Rawat} {et~al.}(2022){Rawat}, {Pandey}, {Joshi}, \&
  {Yadava}}]{rawat2022}
{Rawat}, N., {Pandey}, J.~C., {Joshi}, A., \& {Yadava}, U. 2022, {A step
  towards unveiling the nature of three cataclysmic variables: LS Cam, V902
  Mon, and SWIFT J0746.3-1608}, \mnras, 512, 6054,
  \dodoi{10.1093/mnras/stac844}

\bibitem[{{Shara} {et~al.}(2017){Shara}, {I{\l}kiewicz}, {Miko{\l}ajewska},
  {Pagnotta}, {Bode}, {Crause}, {Drozd}, {Faherty}, {Fuentes-Morales},
  {Grindlay}, {Moffat}, {Pretorius}, {Schmidtobreick}, {Stephenson}, {Tappert},
  \& {Zurek}}]{shara2017}
{Shara}, M.~M., {I{\l}kiewicz}, K., {Miko{\l}ajewska}, J., {et~al.} 2017,
  {Proper-motion age dating of the progeny of Nova Scorpii AD 1437}, \nat, 548,
  558, \dodoi{10.1038/nature23644}

\bibitem[{{Taghizadeh-Popp} {et~al.}(2020){Taghizadeh-Popp}, {Kim}, {Lemson},
  {Medvedev}, {Raddick}, {Szalay}, {Thakar}, {Booker}, {Chhetri}, {Dobos}, \&
  {Rippin}}]{taghizadeh2020}
{Taghizadeh-Popp}, M., {Kim}, J.~W., {Lemson}, G., {et~al.} 2020, {SciServer: A
  science platform for astronomy and beyond}, Astronomy and Computing, 33,
  100412, \dodoi{10.1016/j.ascom.2020.100412}

\bibitem[{{Wada} {et~al.}(2017){Wada}, {Tsujimoto}, {Ebisawa}, \&
  {Hayashi}}]{wada2017}
{Wada}, Q., {Tsujimoto}, M., {Ebisawa}, K., \& {Hayashi}, T. 2017, {A
  systematic X-ray study of the dwarf novae observed with Suzaku}, \pasj, 69,
  10, \dodoi{10.1093/pasj/psw114}

\bibitem[{{Walker}(1954)}]{walker1954}
{Walker}, M.~F. 1954, {Nova DQ Herculis (1934): an Eclipsing Binary with Very
  Short Period}, \pasp, 66, 230, \dodoi{10.1086/126703}

\bibitem[{{Warner}(1986)}]{warner1986}
{Warner}, B. 1986, {Multiple optical orbital sidebands in intermediate
  polars.}, \mnras, 219, 347, \dodoi{10.1093/mnras/219.2.347}

\bibitem[{{Warner} \& {Woudt}(2009)}]{warner2009}
{Warner}, B., \& {Woudt}, P.~A. 2009, {The eclipsing intermediate polar V597
  Pup (Nova Puppis 2007)}, \mnras, 397, 979,
  \dodoi{10.1111/j.1365-2966.2009.15006.x}

\bibitem[{{Witham} {et~al.}(2007){Witham}, {Knigge}, {Aungwerojwit}, {Drew},
  {G{\"a}nsicke}, {Greimel}, {Groot}, {Roelofs}, {Steeghs}, \&
  {Woudt}}]{witham2007}
{Witham}, A.~R., {Knigge}, C., {Aungwerojwit}, A., {et~al.} 2007, {Newly
  discovered cataclysmic variables from the INT/WFC photometric
  H{\ensuremath{\alpha}} survey of the northern Galactic plane}, \mnras, 382,
  1158, \dodoi{10.1111/j.1365-2966.2007.12426.x}

\bibitem[{{Worpel} {et~al.}(2018){Worpel}, {Schwope}, {Traulsen}, {Mukai}, \&
  {Ok}}]{worpel2018}
{Worpel}, H., {Schwope}, A.~D., {Traulsen}, I., {Mukai}, K., \& {Ok}, S. 2018,
  {V902 Monocerotis: A likely disc-accreting intermediate polar}, \aap, 617,
  A52, \dodoi{10.1051/0004-6361/201833472}

\end{thebibliography}
\bibliographystyle{psj}
\end{document}